# Production of high-quality plasma discharges via real-time control of plasma current ramp-up using neutral gas injection in Aditya-U tokamak.

Suman Dolui[1,2] , Praveenlal EV[1], Kaushlender Singh[1,2], J. Ghosh[1,2], R L Tanna[1], K A Jadeja[1] , Ankit Patel[1] , A.Kundu[1], Rohit Kumar[1], Suman Aich[1] , Harshita Raj[1,2], K M Patel[1], M B Chowdhuri[1], S.Purohit[1], P. K. Chattopadhyay[1,2], A.Sen[1,2], R.Pal[4]

*[1] Institute for Plasma Research, Gandhinagar 382 428*
*[2] Homi Bhabha National Institute (HBNI), Mumbai 400 085*

**Abstract:** Robust control of plasma current ramp-up is an absolute necessity, as an efficient and uncontaminated plasma current ramp-up is essential for achieving prolonged, high-pressure tokamak plasma discharges. In conventional tokamaks with Ohmic breakdown, the plasma current ramp-up is achieved primarily with pre-fixed temporal profiles of the applied toroidal electric field and the equilibrium magnetic field ($B_V$). The pre-fixed temporal profiles of these fields are often insufficient to maintain a successful plasma current ramp-up, as several unquantified dynamical variables, such as the condition of the vessel wall and plasma-facing components that influence the plasma current rise. Fuel gas injection in an appropriate quantity at a suitable time during the current ramp-up is therefore used to control the plasma current rise rate, ensuring successful plasma current start-up in ADITYA-U. The gas injection time and gas quantity are controlled based on real-time measurement of plasma current rise rate using a Digital Signal Processor (DSP) controller. This special control scheme is capable of achieving the plasma current to rise nearly at the desired rate, resulting in a successful start-up and a stable plasma discharge.

## I. Introduction:

In conventional tokamaks plasma is typically initiated through the central solenoid–induced breakdown process[1,2], where the time evolution of the plasma current ($I_p$) is governed by the externally applied electric field, induced by the transformer primary current, or equivalently, the central solenoid current waveform. In order to sustain a desired plasma current profile and ensure stable plasma discharge, an appropriate magnitude and decay index of the $B_V$ are required to maintain the plasma column in equilibrium against electromagnetic hoop forces[3]. The overall plasma discharge can be divided into several phases, each posing distinct operational challenges. Following the successful completion of the breakdown and burn-through phases[2], the plasma current undergoes a ramp up until a quasi-stationary flat-top is achieved. Among these stages, the ramp-up phase is especially crucial, since it is susceptible to failures in plasma column stability and position control. During the ramp-up phase, the dynamic evolution of the plasma current often deviates from the pre-planned profile. The time evolution of the plasma current is significantly influenced by the condition of the vessel and plasma-facing components prior to each plasma discharge. Plasma–wall interactions during the previous discharge may introduce unquantified impurities that affect the current rise rate in the subsequent discharge. Consequently, the dynamically evolving plasma current cannot always be supported by the pre-programmed vertical equilibrium magnetic field ($B_V$). Even small mismatches between the required and programmed $B_V$ can lead to loss of horizontal position control of the plasma column. This can result in enhanced

plasma–wall interactions, impurity influx, and a consequent deterioration in discharge quality, often culminating in plasma disruptions. Real-time control of either the current-driving electric field or the vertical equilibrium magnetic field ($B_V$) is required to achieve successful plasma discharges. In conventional Ohmic breakdown with no or limited pre-ionization, real-time control of the applied toroidal electric field remains challenging. In such cases, real-time control of $B_V$ alone is adopted in most tokamaks. However, the rapid response required for feedback-based plasma position control is limited by the large time constant ($\tau$) of the primary $B_V$ coils compared to the characteristic time scale of plasma column motion. The ADITYA-U Tokamak is a conventional tokamak with Ohmic breakdown, whose operation has been described previously. In ADITYA-U, the time constant of the primary $B_V$ coils is about 20 ms, which is significantly larger than the characteristic time scale ($\sim 1\ ms$) of plasma column motion. To overcome this limitation, auxiliary fast feedback (FFB) coils with a much lower time constant ($\tau \sim 1\ ms$) are employed to provide real-time feedback stabilization. The ramp-up phase is therefore highly sensitive. A slower-than-expected rise of $I_p$ can be compensated using the auxiliary FFB coils[4] with low inductance, whereas a faster-than-expected rise may exceed the control capability of the power supplies for both the main vertical field (VFPS) and the auxiliary coils (FFBPS). When the plasma current evolves with an undesirably high ramp rate ($dI_p/dt$) exceeding$\sim 6\ kA/ms$, the combined VFPS and FFBPS systems become insufficient to sustain the required equilibrium field for stable plasma operation[3]. Even when a high current rise rate ($\sim 6\ kA/ms$) is supported by real-time control of $B_V$ to maintain a stable plasma column, MHD instabilities may still be triggered solely due to the large current rise rate. Beyond issues of position control, such rapid current ramping can destabilize the current density profile and trigger tearing modes or other MHD-driven disruptions. These instabilities are closely linked to the interplay between electron temperature ($T_e$) and density evolution. As $T_e$ rises, the penetration of the plasma current decreases, and this can result in tearing mode instabilities during plasma ramp-up[5]. Hence, the plasma condition with larger resistive current diffusion time cannot sustain the high $I_p$ ramp rate as the plasma disruption occurs due to MHD instabilities. Hence, during the ramp-up phase, real-time control of the $I_p$ along with the $B_V$ is essential to achieve a stable plasma discharge and avoid disruptions caused by MHD instabilities triggered either by plasma–wall interactions or by an excessively large current rise rate. The most effective way to accomplish this is to maintain control over the plasma current ramp rate. Keeping the plasma current rise rate within a safe limit can help ensure stable operation. In the ADITYA-U Tokamak, a novel active $I_p$ control strategy has been developed. This method involves the injection of short, controlled fuel gas pulses whenever the current ramp rate ($dI_p/dt$) exceeds a critical threshold, ensuring that it does not surpass the desired rise rate. A proof-of-principle experiment demonstrated that such gas injection effectively reduces the current ramp rate, thereby preventing potential disruption. The real-time implementation of this approach relies on dedicated hardware, including a DSP-based controller capable of monitoring the plasma current ramp rate and generating a TTL pulse for actuating the gas injection system once the preset threshold is crossed. This study describes the experimental implementation of this disruption mitigation strategy in ADITYA-U. Section II describes the experimental setup and real-time gas pulse injection mechanism. Section III presents the results of the proof-of-principle experiments, while Section IV discusses the implications of the findings and summarizes the key outcomes.

## II. Experimental setup

### a. Aditya-U plasma operation regime:

The experiments were performed on the ADITYA-U tokamak[6], which is a medium-sized, Ohmically heated, limiter tokamak with $R_0$=0.75 m, a=0.25 m, to investigate the effect of controlled gas injection during the plasma current ramp-up phase. The experiments cover a wide range of plasma parameters: $I_P \sim 80 - 200\ kA$, line-averaged density$(n_e) \sim 1 - 4 \times 10^{19} m^{-3}$, $B_T \sim 0.9 - 1.44\ T$, core electron temperature $(T_{e0}) \sim 250\ ev - 500\ eV$. An ultra-high vacuum (UHV) of $\sim 6 - 9 \times 10^{-9}$ Torr is sustained, and the $H_2$ plasmas are produced at pre-filled pressure $\sim 1 - 2 \times 10^{-4}$ Torr. Plasma breakdown and the desired time profile of plasma current at the subsequent stages can be obtained by the appropriate shaping of the central solenoid current. A schematic overview of the diagnostic and control systems employed in this study is presented below.

### b. Relevant diagnostics for the experiment:

Hydrogen gas puffs were injected into the vacuum vessel using a piezo-electric valve installed at one of the bottom ports of the machine. Each pulse of width 1 ms corresponds to approximately $\sim 10^{17}$ neutral particles. The injection was triggered in real time based on the monitored plasma current ramp rate. The line-averaged plasma density was monitored using a 100 GHz heterodyne microwave interferometry system[7], which provided real-time chord-averaged density data during the discharge. A soft X-ray (SXR) tomography[8] camera, equipped with AXUV (absolute extreme ultraviolet) photodiodes, was employed to probe the evolution of the core plasma region. Complementary spectroscopic diagnostics[9] were used to characterize impurity influx and plasma–surface interactions. Temporal evolutions of visible line emissions from hydrogen$(H_\alpha)$, carbon$(C^{2+})$, and oxygen $(O^{1+})$,) were monitored using photo-multiplier tubes (PMTs) coupled with interference filters. These chord-averaged measurements were obtained from a mid-plane viewing port aligned along a radial line of sight from the outboard mid-plane to the inner limiter. Plasma current $(I_p)$ was measured using Rogowski coils. The horizontal position of the plasma column was determined using a calibrated cosine coil diagnostic mounted outside the vacuum vessel. To cross-check these measurements, fast camera imaging was used to record the plasma column dynamics. In addition to the cosine coil diagnostic[10], using spectroscopic diagnostics, edge emissions from the inboard (high toroidal field side) and outboard (low toroidal field side) plasma boundary areas were monitored in order to approximate the horizontal plasma position.

### c. A special hardware to control plasma current rise rate:

A special DSP controller based hardware (DCBH) has been developed to monitor the plasma current rise rate during the ramp up phase. The hardware integrated signal of a rogowski coil is multiplied by a calibration factor to determine the exact value of the plasma current. The output of the rogowski integrator is fed to the DCBH. Any given external trigger, generally generated by the peak loop voltage during the breakdown, makes the DCBH start functioning. A suitable time delay after the external trigger can be set in order to start the $I_p$ slope measurement precisely from the beginning of the ramp-up phase. The time interval $(\Delta t)$, to determine the current rise rate $\left(\frac{\Delta I_p}{\Delta t}\right)$, can be varied at any value. However, in our experiment at every 5 ms time interval, the slope of the plasma current was measured. The DCBH can generate a TTL pulse while the value of the current slope crosses a preset threshold value. The TTL pulse externally triggers a function generator (FG) at the following stage. The FG was used to generate square pulses of different widths and then pulse amplitude is modified by a voltage amplifier to a certain value required to activate the piezo electric valve. The amount of injected gas depends on the width and amplitude of the pulse applied at the piezo electric gas valve, installed at bottom port of the machine. The block diagram of the entire system described above is shown in Fig.1.

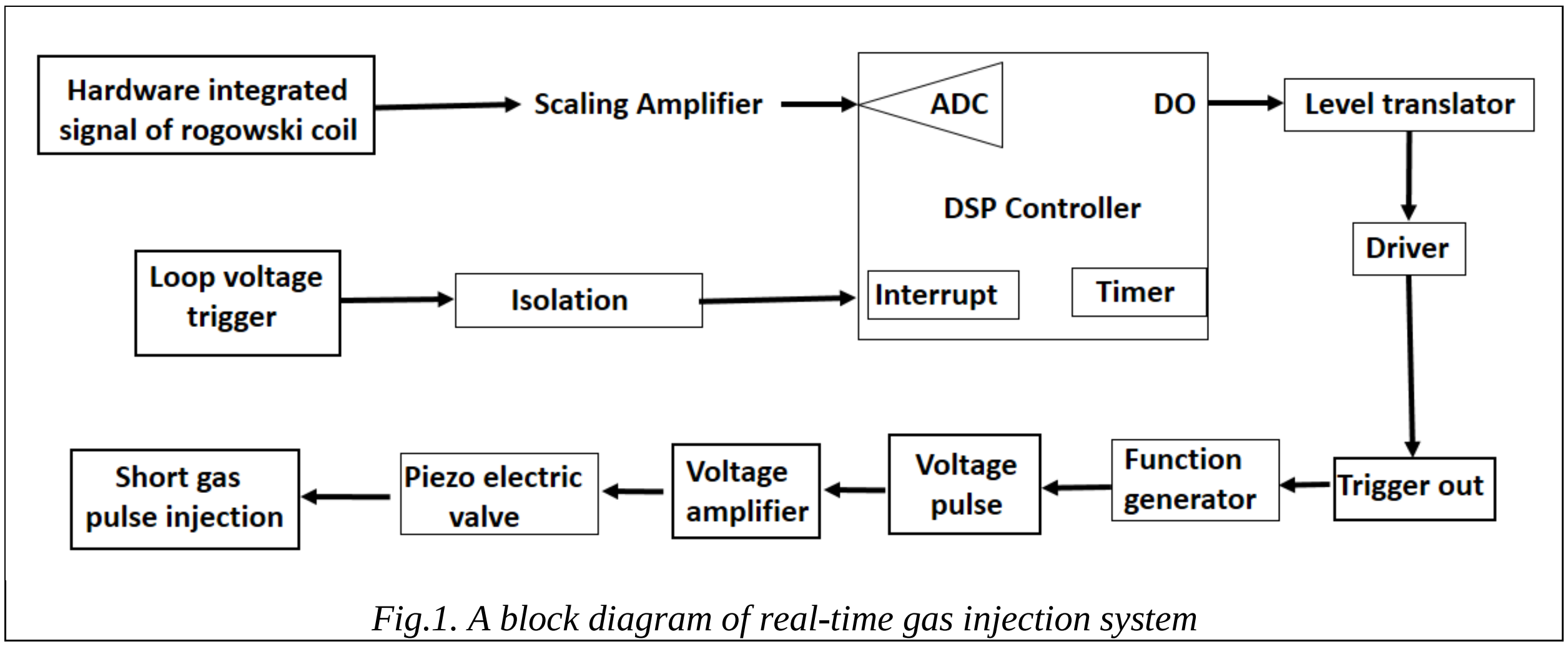

*Fig.1. A block diagram of real-time gas injection system*

## III. Experimental results

### a. Functionality test of the hardware

An experiment was carried out to verify the operational reliability and performance of the DSP controller–based hardware (DCBH). As illustrated in Fig. 2, the system was tested in two separate plasma discharges. The red dashed vertical line indicates the loop voltage trigger, which activates the DCBH. A delay of 5 ms was introduced to ensure that slope measurements began after successful breakdown, during the plasma current ramp-up phase. The green dashed vertical lines represent 5 ms intervals over which the current slope was evaluated. The DCBH continuously monitors the integrated Rogowski coil signal and calculates the voltage difference over successive 5 ms intervals. When this difference exceeds a threshold of 0.3 V, a TTL pulse is generated. The relationship between plasma current and integrated Rogowski voltage is expressed as:

$$I_p(t)(kA) = V(t)_{integrated} \times 53.24$$

$$\frac{dI_p}{dt} = 53.24 \times \frac{dV_{int}}{dt}$$

$$\frac{dI_p}{dt} = 53.24 \times \frac{0.3}{5} \sim 3.2\ kA/ms$$

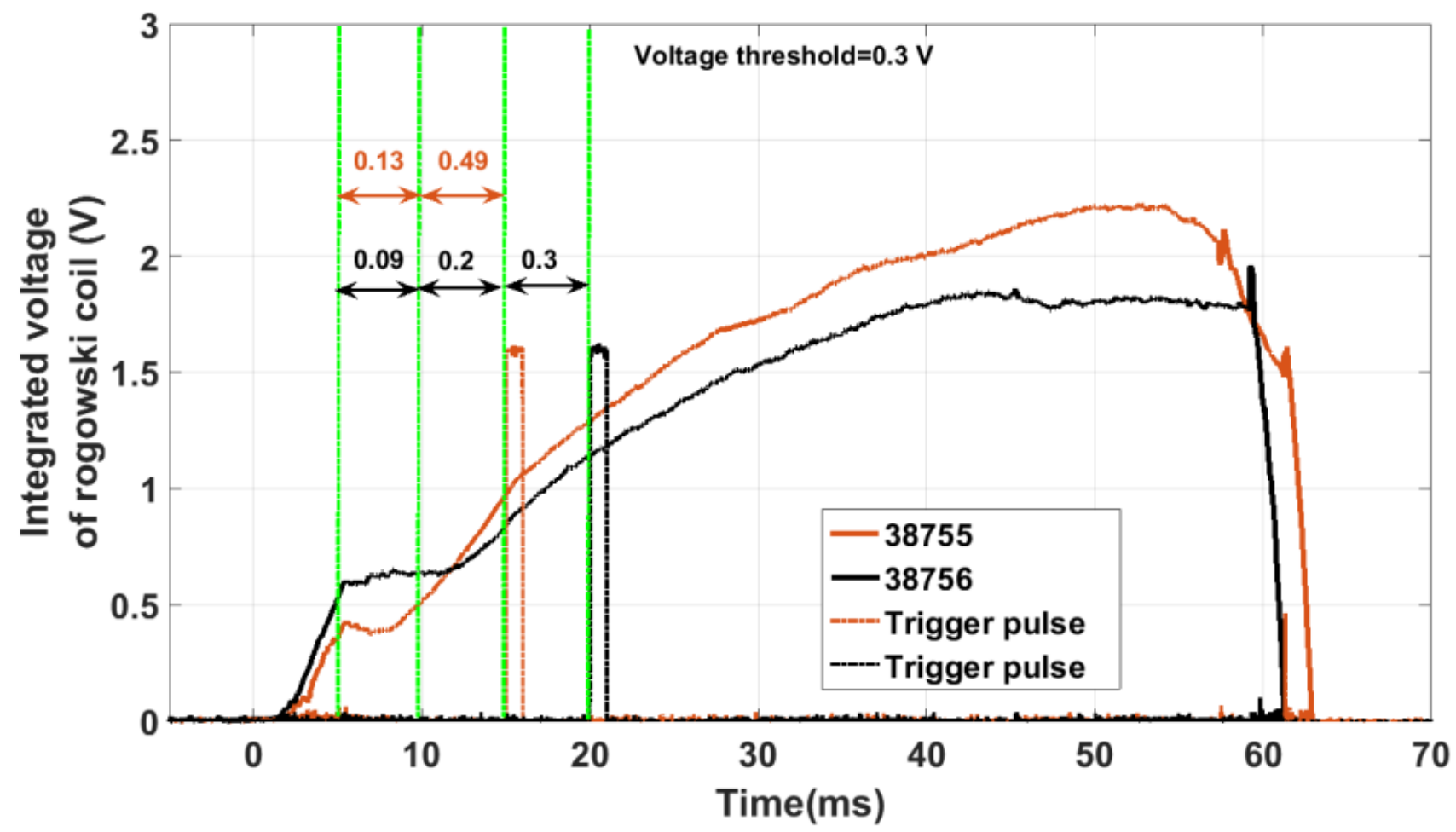

*Fig.2. Time evolution of hardware integrated signal of rogowski coil for two different plasma discharges #38755 and #38756.*

Thus, a voltage difference of 0.3 V over 5 ms corresponds to a current rise rate of approximately 3.2 kA/ms. For discharge #38755 (brown trace in Fig. 2), the voltage difference during the first interval (5–10 ms) was 0.13 V, below the threshold. In the following interval (10–15 ms), the difference increased to 0.49 V, exceeding the threshold and resulting in a TTL pulse at 15 ms (brown dotted pulse). In contrast, for discharge #38756 (black trace), the ramp-up rate was slower, and the threshold condition was satisfied later, producing a TTL pulse only at 20 ms. These results confirm that the DCBH accurately tracks the plasma current slope in real time and generates trigger signals at the appropriate thresholds, demonstrating reliable functionality for ramp-rate monitoring. In this experiment the gas injection system was remained inactive. Only the functionality of the DCBH system was tested.

### b. Plasma control scheme in Aditya-U

The plasma discharge production in Aditya-U relies on three critical preset parameters: prefill gas pressure, central solenoid (CS) current shaping, and a preprogrammed vertical equilibrium magnetic field ($B_V$).The prefill gas pressure primarily influences the time required for breakdown completion at a given toroidal electric field. Following plasma initiation, the time-dependent variation of CS current determines the induced loop voltage at different stages of the discharge, thereby governing the evolution of the plasma current waveform. The vertical equilibrium magnetic field is essential for maintaining radial force balance of the plasma column. It is expressed as:

$$B_V = -\frac{\mu_0 I_p}{4\pi R_0}\left[\ln\left(\frac{8R_0}{a}\right) + \frac{l_i}{2} + \beta_p - \frac{3}{2}\right]$$

Where $I_p$,$R_0$, a, $\beta_p$ and $l_i$ denote the plasma current, major radius, minor radius, plasma poloidal beta, and internal inductance, respectively. For typical Aditya-U parameters, the required vertical field approximately follows the relation: $B_V \sim 4I_p$.

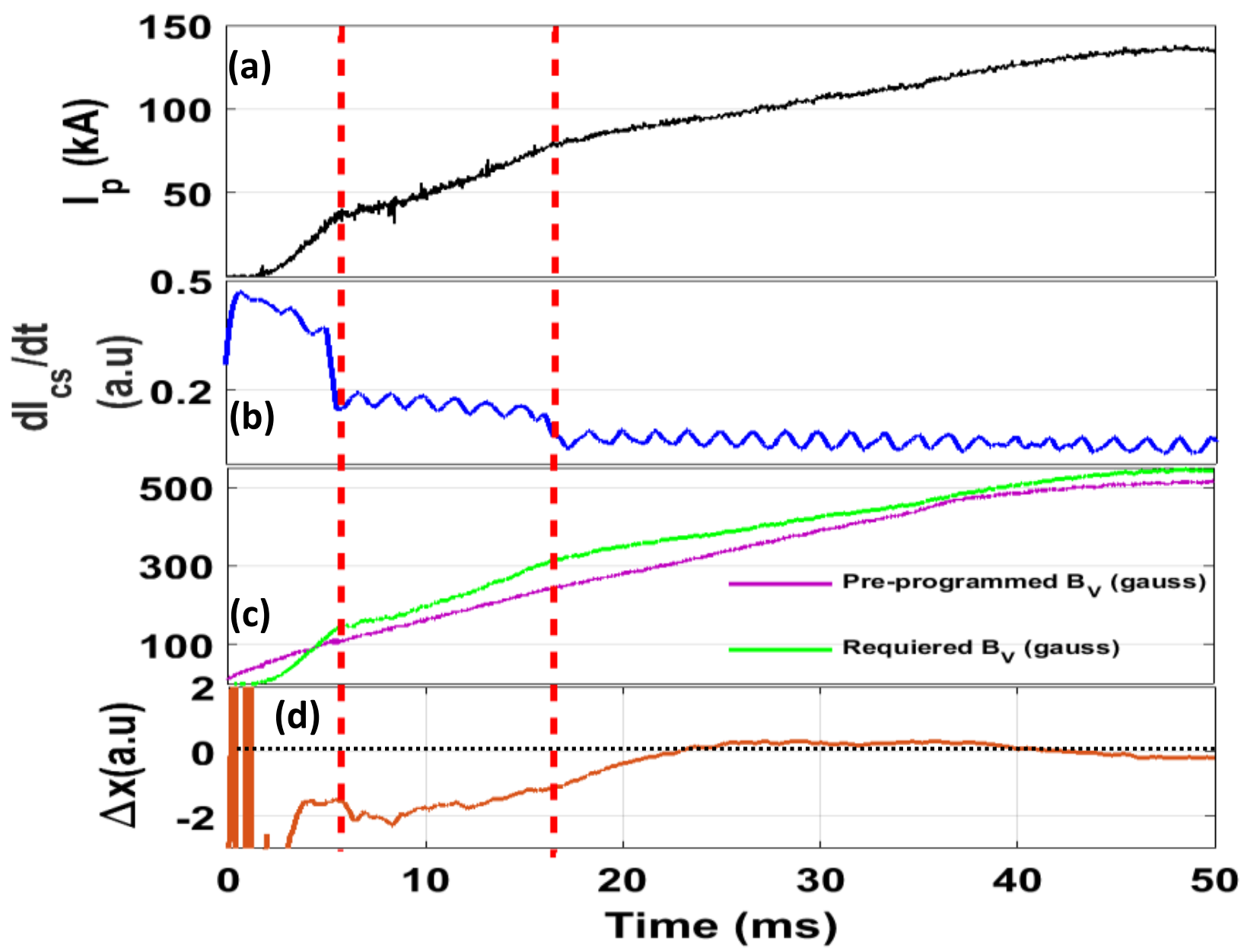

*Fig.3. Time evolution of Aditya-U discharge parameters (a) plasma current (b) Current variation rate in central solenoid coil (c) Comparison between pre-programmed and required vertical equilibrium magnetic field ($B_V$) (d) horizontal plasma position measured by the cosine coil.*

Figure 3a shows that the plasma current undergoes noticeable transitions (highlighted by vertical red dashed lines) whenever there is a sharp change in the CS current variation rate (Fig. 3b). During the time interval of 5–25 ms, the plasma column gradually shifts outward ($\Delta x > 0$) from an initially compressed inward state, as indicated by the time evolution of the cosine signal (Fig. 3d). Until the complete plasma formation (initial 5 ms), the preset $B_V$ is always higher than the required equilibrium magnetic field. In the plasma formation phase (first 5 ms), the preset $B_V$ is always higher than the required equilibrium value. As a result, the plasma column remains naturally displaced toward the inboard side. Following complete plasma formation, the carefully optimized ramp rate ensures that the difference between the preset $B_V$ and the required equilibrium $B_V$ (Fig. 3c) allows the plasma column to gradually shift toward the vessel center (horizontal black dotted line in Fig. 3d). This discharge thus exemplifies a scenario where plasma position control is effectively achieved during the early transient phase, even before reaching the steady flat-top regime, and notably, without any assistance from a real-time feedback control system.

### c. Impact of gas pulse injection on plasma control in real time

The highly unpredictable dynamic nature of experimental scenario often causes the plasma current evolution to deviate from the intended waveform. One of the major sources of this unpredictability is the pre-discharge vacuum vessel condition, which strongly influences the breakdown and burn-through phases and, in turn, the current ramp-up behavior. Since these conditions cannot be fully incorporated into pre-planned waveforms, uncontrolled deviations frequently arise during experiments. As shown in Fig. 3, when the current ramp rate remained below 4 kA/ms, stable control was maintained. However, exceeding this value led to loss of equilibrium support from the vertical field ($B_V$), outward plasma displacement, and intense boundary interactions.

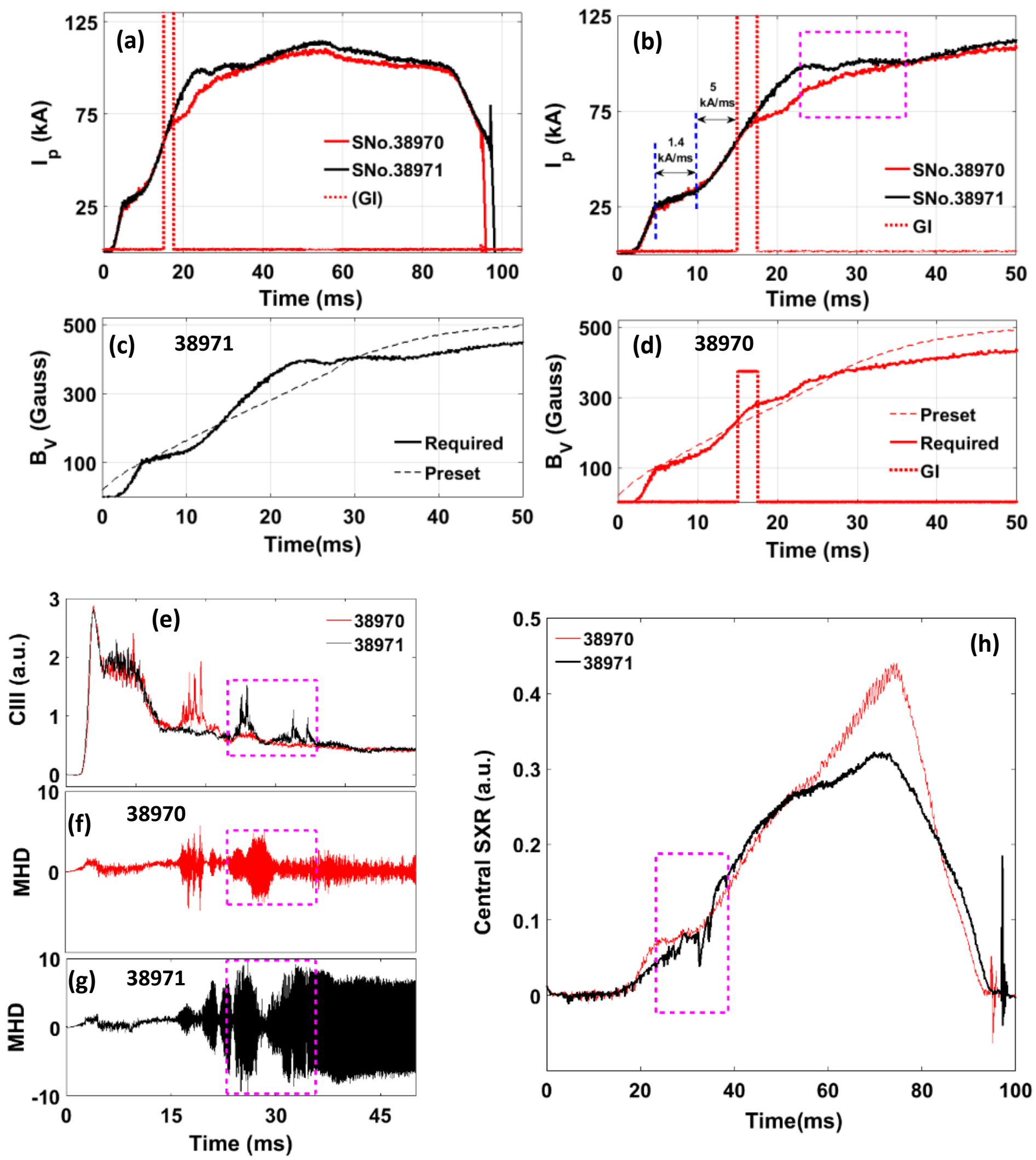

*Fig.4. Time traces of (a) plasma current for discharges #38970 and #38971, and (b) its temporally zoomed-in view over 0–50 ms. Comparison of the pre-programmed and required vertical equilibrium magnetic field ($B_V$) for (c) discharge #38971 and (d) discharge #38970. (e) $C^{2+}$Line emission intensity for the corresponding discharges. (f, g) MHD activity in discharges #38970 and #38971, respectively. (h) Central chord-averaged SXR intensity comparison between the two discharges.*

To evaluate real-time control mechanism, two consecutive plasma discharges with identical pre-discharge settings were compared (Fig. 4a). In discharge #38970, the gas injection (GI) system was activated, while in #38971 it was deactivated. As illustrated in Fig. 4b, when the plasma current ramp rate exceeded the 4 kA/ms threshold and reached 5 kA/ms, a 2.5 ms gas puff (vertical red pulse) was triggered. This intervention reduced the ramp rate to below 4 kA/ms. In contrast, the uncontrolled discharge (black trace) exhibited continued rapid current rise, followed by a slowdown due to plasma–boundary interactions. Figs. 4c–d. show without gas injection, the equilibrium field requirement deviated significantly from the preset profile, while active control maintained close agreement. The dashed magenta boxes (23–36 ms) mark the period of strongest plasma–surface interaction, evidenced by spikes in $C^{2+}$ line emission intensity (Fig. 4e) and accompanied by large MHD activity (Fig. 4g). Such events are associated with edge cooling, impurity radiation, and subsequent degradation of discharge quality, often culminating in major disruptions. Finally, the comparison of central chord-averaged soft X-ray (SXR) intensity (Fig. 4h) illustrates the effect of real-time control. The uncontrolled plasma (black trace) showed multiple spikes, highlighted by the box, characteristic of thermal quenches, while the actively controlled plasma (red trace) exhibited a peaked SXR profile with sawtooth oscillations[11], indicative of a high-pressure, stable discharge. These results clearly demonstrate that gas pulse injection, triggered in real time, mitigates excessive ramp rates and prevents disruption-prone plasma–wall interactions, thereby enhancing both stability and discharge performance.

## IV. Summary and future work

Plasma current rise rate in tokamaks is governed by the relation[12]

$$\frac{dI_p}{dt} = \frac{1}{L_p}\left(V_{loop} - I_p R - \frac{I_p dl_i}{dt}\right)$$

Where $L_p$,$V_{loop}$, $l_i$ , R represent the total inductance, loop voltage, internal inductance, and plasma resistance, respectively. Plasma resistance is determined by the plasma temperature and the impurity content within the vessel. The internal inductance, in contrast, is governed by the radial distribution of the current density. Both R and $l_i$ evolve dynamically in time and exhibit a strong interdependence which play a central role in shaping the plasma current ramp-up process. In large, advanced tokamaks, closed-loop feedback control using high-performance power supplies provides the primary mechanism for regulating plasma current[13–15]. However, our study demonstrates that real-time neutral gas injection offers a complementary means of controlling the current rise rate. Experimental observations reveal that following a gas pulse, plasma density increases concurrently with impurity line emissions (notably $C^{2+}, O^{1+}$) and enhanced $H_\alpha$ radiation. These signatures indicate energy loss via line radiation and ionization processes, which in turn reduce plasma temperature and increase resistivity. The higher resistivity shortens the resistive diffusion time, enabling deeper current penetration and leading to a less hollow current density profile. Consequently, the internal inductance increases with time, collectively suppressing the current rise rate[16]. The results further suggest that the gas injection scheme allows the rise rate to remain below the maximum threshold manageable by advanced feedback control systems, making it a valuable auxiliary tool for plasma current regulation. Since every active control system has an intrinsic upper limit on how fast it can regulate current ramp-up, the addition of a gas injection scheme enhances operational robustness by ensuring the ramp rate remains within manageable limits. For practical implementation, optimization of gas quantity is essential as excessive injection could trigger disruptions similar to massive gas injection events. Current studies are examining how different gas amounts affect the current rise rate, with the goal of developing an injection strategy that is both efficient and reliable. In summary, the neutral gas injection scheme, when integrated with existing closed-loop systems, emerges as a promising auxiliary control strategy for regulating plasma current ramp-up in tokamaks.